# An Analysis of Technical Debt Management Through Resources Allocation Policies in Software Maintenance Process

**Eduardo Ferreira Franco, Joaquim Rocha dos Santos, Hamilton Carvalho, Marcelo Ramos Martins and Kechi Hirama**

Escola Politécnica of University of São Paulo
Av. Prof. Luciano Gualberto, travessa 3 nº 158, CEP 05508-900, São Paulo, SP, Brazil
Tel. / Facsimile: +55 11 3091-5583 / +55 11 3091-5294
E-mail: {eduardo.franco, jrsantos, hccarvalho, mrmartin, kechi.hirama}@usp.br

*Abstract — This paper presents an analysis of technical debt management through resources allocation policies in software maintenance process during its operation to demonstrate how different strategies leads to the emergence of different behaviors along the evolution path. To achieve this objective, this work used the System Dynamic approach for building a computational simulation model based on extensive literature review and secondary data. Most of the works that applied the System Dynamics on software projects research, focused on initial phases of its life cycle, leaving a gap to be explored regarding the long-term behaviors of the operation and maintenance phases. The results demonstrated that the excessive focus on the perfective maintenance activities could be more costly than performing regular preventive maintenance to reduce the technical debt incurred, ending up with fewer functionalities deployed, higher backlog, lower productivity, lower maintainability and higher technical principal.*

## 1 INTRODUCTION

There is an understanding that software deployment projects are broader than just placing artifacts in operation. Their introduction alters the structure and culture of an organization, in addition, they change the way people think and work (Orlikowski & Robey, 1991). The complexity associated with such initiatives is characterized by involving interactions between technological components, people, information and organizational issues, which creates a dynamically complex environment, containing feedback loops, accumulations and delays between causes and effects, which has emergent behaviors and requires non-trivial and non-intuitive solutions (Georgantzas & Katsamakas, 2008).

Defining success of those initiatives is also a non-trivial activity, and consequently, there is no consensus in the research community about how to define and measure it. From the 1980s, many were proposed for evaluating its success (Franco, So, Maximiano, & Hirama, 2015).

Conversely, some authors suggest that a software initiative can only be considered a failure when its development or operation is canceled. Based on this criterion of failure, software products are similar to natural systems, where the observed behaviors are explained in terms of survival goals (Sauer, 1993). Its survival is achieved through the supply of resources that support the continuity of its operation. Thus, it cannot be considered a failure while still operating and attracting resources (Yeo, 2002).

The comparison of software products success to the survival of natural systems is strengthened by the fact that the majority of the resources allocated to those initiatives, approximately 80%, occurs during the operating and maintenance phase, which is considered to be any changes made to the system after the beginning of its operation (INCOSE, 2015).

This imbalance caught the attention of the scientific community and, from the 1970s; researchers began to investigate the possible causes for the demand for constant investments (M. Lehman, 1980). These investigations led to a new field of research in software engineering area, called "Software Evolution" and the consolidation of the laws of evolution that describe abstractions of observed behaviors through statistical models (M. Lehman & Ramil, 2006). As a software system ages, it needs to be constantly modified and expanded to continue to meet the business needs and objectives. These changes, due to violations of architecture and coding good practices, lead to a decrease in system quality over time.

The metaphor "technical debt" was created to describe the liability accumulated by the decisions that happened in the past, intentional or not, to deliver software systems containing sub-optimal code quality to achieve business objectives (Cunningham, 1993). This technical debt incurs in an increase in the cost associated with maintenance activities, an increase in unresolved errors, reduces its modifiability to meet current and future business needs, and therefore, reduces users' satisfaction in the long-term evaluation.

In this way, Parnas (1994, p. 2) reinforced the importance of the maintenance phase stating that:

> *"A sign that the Software Engineering profession has matured will be that we lose our preoccupation with the first release and focus on the long-term health of our products. Researchers and practitioners must change their perception of the problems of software development. Only then will Software Engineering deserve to be called Engineering."*

The research question that guided this work was to understand, from the maintenance process perspective, why even after the beginning of its operation, software products demands continuous investments to maintain sufficient levels of its quality attributes (functionality and maintainability) and how different resources allocation policies affect the behavior of those attributes throughout its evolution?

The main contribution is to propose and develop a simulation model that permits to expand the current knowledge by exploring and evaluating the impact that different maintenance resources allocation policies have on software adaptability and evolution capabilities and quality attributes related to functionality, maintainability and costs during the phases of operation, maintenance, and deactivation. The model will also support the investigation and evaluation of maintenance policies that sustain adequate levels of quality attributes throughout the life cycle of these systems and optimize the tradeoff analysis of the compromise between technical debt accumulation, maintenance cost and software quality attributes.

The methodology applied is the Systems Dynamics approach. It was developed in the 50's, by Jay Forrester (Forrester, 1961), to study complex business problems and was later expanded to study problems associated with sustainability of population growth, global warming, and many others. This approach consists in an iterative process to define a dynamic hypothesis, develop a formal model, test it, and to formulate and evaluate different intervention policies (Sterman, 2000).

This work is organized into six sections. Section 2 presents a summary of related works previously published in the context being explored for the formulation of the research context. Section 3 presents the problem articulation and the dynamic hypothesis adopted. Section 4 presents the model formulation, followed by discussions of the preliminary results obtained from the simulation of different maintenance investments strategies in Section 5. Finally, Section 6 presents the conclusions, the limitations of the current work and suggestions for future works.

## 2 RELATED WORKS

The interest of researchers and practitioners in process modeling and simulating has grown. It has been perceived as an approach: capable of analyzing complex business context; to support policy design and evaluation; to perform tests and experiments for evaluating scenarios that would often be economically unfeasible to explore in the real world. Although modeling and simulation techniques have been employed widely in a variety of disciplines for a long time, only recently it has been applied to software development and process improvement areas (Kellner, Madachy, & Raffo, 1999).

This growth of interests can be rooted to the increased demand for better results in the last years due to the growing pressure for faster deliveries, lower costs, scope flexibility, system interconnectivity and business dependence on information systems for operating its daily activities. These demands require changes in software development processes and in the organizations adopting those systems, which requires significant investments that are complex to evaluate. How is it possible to understand and anticipate the impacts these changes present? The "Software Process Simulation and Modeling" (SPSM) is one area of research that has sought to address this issue and have brought contributions to better evaluate scenarios and predict potential impacts of proposed software process improvements (Kellner et al., 1999; Ruiz, Ramos, & Toro, 2004).

There are several approaches for building and simulating models (Petri nets, agent-based, Monte Carlo, Bayesian networks etc.); however, a literature review exploring studies on the application of simulation in the software industry, published between 1998 and 2012, indicates that the predominant approach applied is the System Dynamics, corresponding to approximately 37% of the works (Ali, Petersen, & Wohlin, 2014).

In the mid-80s, works applying the System Dynamics approach to study the dynamics associated with software projects began to emerge (Abdel-Hamid & Madnick, 1982; Abdel-Hamid, 1984), which proliferated in the 90´s (Abdel-Hamid & Madnick, 1991; Kellner et al., 1999; Lin, Abdel-Hamid, & Sherif, 1997; Waeselynck & Pfahl, 1994; Wernick & Lehman, 1999).

The existing works focus on software development stages or part of the development cycles. There is a gap of works that explore the implementation and post implementation phases of software products initiatives, involving the interaction with end users, maintenance and the long-term evaluation of the return on investments.

Recently, some works were published dealing with the operation, maintenance, and especially software sustainment, regarding military system of systems based on software (Ferguson, Phillips, & Sheard, 2014; Sheard, Ferguson, Phillips, & Moore, 2014; Sheard, Ferguson, Moore, & Phillips, 2015). Although dealing with long-term maintenance effects, those works did not distinguished maintenance activities types nor did the deal with software product quality attributes, such as maintainability and technical debt accumulation.

## 3 THEORY BACKGROUND

### 3.1 Software Maintenance

The software inventory owned by a company represents a significant share of its assets (Wiederhold, 2006). Hence, companies have a vital interest in preserving or extending the value of their software repositories. To achieve this, companies need to counter the gradual decay that a software system's value is known to undergo (M. M. Lehman, Ramil, Wernick, Perry, & Turski, 1997; Parnas, 1994) by continuously adapting the software product to changing requirements.

The continuous growth of maintenance costs over the years is partially explained by the growth of software products lifetime still in use. As Table 1 shows, in fifteen years (from 1990 to 2005), the average age of Management Information Systems (MIS) doubled, and Web applications, which appeared near 1995, more than tripled in ten years. With long periods of operation, software products grow due to new requirements added to adapt to emergent scenarios and environments not previously planned. As a result, of these continuous changes, software products maintainability decreases which in turn, increases its maintenance costs.

Table 1 – Average age in years of software application still in use.

| Software application | 1990 | 1995 | 2005 |
|---|---|---|---|
| End-user | 1.50 | 2.00 | 2.00 |
| Web | - | 1.50 | 5.00 |
| MIS | 10.00 | 15.00 | 20.00 |
| Outsourced | 5.00 | 7.00 | 9.00 |
| Systems | 5.50 | 8.00 | 12.00 |
| Commercial | 2.00 | 2.50 | 3.80 |
| Military | 12.00 | 16.00 | 23.00 |
| Average | 5.14 | 7.43 | 10.69 |

Source: Adapted from (Jones, 2008).

There are numerous definitions of software maintenance that typically define it as the process of modifying software after its initial delivery. In this work, we use the term "software maintenance" as described by the standard ISO/IEC 14764/2006: "the totality of activities required to provide cost-effective support to a software system". Moreover, the maintenance activities can be classified in four categories (Deißenböck, 2009):

a) *Perfective*: change existing functionality or add new functionality to a software system, e.g. in response to changes to business processes or new user requests;

b) *Preventive*: prepare a system for prospective changes. This type of maintenance is performed to enhance the efficiency of future maintenance tasks;

c) *Corrective*: correct identified faults in a software system;

d) *Adaptive*: adapt a system to a changing environment, e.g. by changes in base technology like operating systems.

In this work, we analyze the resource allocation in the first two categories of maintenance activities (perfective and preventive) during the software product operation and maintenance. This resource allocation constitutes the trade-off analysis of focusing on delivering business functionality versus keeping the maintainability of the software product.

*3.2 Technical Debt*

The metaphor "technical debt" was coined by Cunningham (1993) and refers to the long-term costs associated with the shortcuts taken by developers, during the development and maintenance of software products, to deliver short-benefits for the business. Despite the simple and intuitive definition, the metaphor has been used indiscriminately to describe any type of impediment, friction and obstacle in the sale, development, deployment, maintenance and evolution of software-based systems, which have weakened and diluted the meaning of metaphor (Kruchten, Nord, & Ozkaya, 2012).

> *"Although immature code may work fine and be completely acceptable to the customer, excess quantities will make a program unmasterable, leading to extreme specialization of programmers and finally an inflexible product.* […] *Shipping first time code is like going into debt. A little debt speeds development so long as it is paid back promptly with a rewrite.* [...] *The danger occurs when* *the debt is not repaid. Every minute spent on not-quite-right code counts as interest on that debt.*" (Cunningham, 1993, p. 30)

The technical debt metaphor consists of a comprehensive set of constructs that help in communicating the costs and risks related to low structural quality of a software program (Curtis, Sappidi, & Szynkarski, 2012):

- *Should-fix violations* are violations of good architectural or coding practices known to have an unacceptable probability of contributing to high costs of ownership, such as excessive effort to implement changes;

- *Principal* is the cost of remediating should-fix violations in production code;

- *Interest* is the continuing costs attributable to should-fix violations in production code that have not been remediated, such as greater maintenance hours;

- *Technical debt* is the future costs attributable to known violations in production code that should be fixed, which includes both principal and interest.

There is no consensus as to what types and which levels of the violations of the software quality attributes that can be classified as a technical debt, and there is no established limits for using the metaphor. In this way, a systematic literature review was conducted to evaluate the use of the metaphor that identified ten categories of technical debt: requirements, architecture, design, code, test, build, documentation, infrastructure, versioning and defects (Li, Avgeriou, & Liang, 2015).

Technical debt can be beneficial or detrimental to the software product operation and maintenance management. Debts that are intentionally incurred to obtain short-term benefits can be positive if the associated costs are kept visible and under control (Allman, 2012). However, they may occur unintentionally and not perceived by those involved. If kept invisible and unresolved, technical debt can accumulate and may pose risks to maintenance activities and long-term evolution (Li et al., 2015).

## 4 DYNAMIC HYPOTHESIS

As stated previously, this work aims to investigate why even after the beginning of its operation; software products demand continuous investments to remain useful, satisfying its users and meeting business needs and how different resource allocation policies affect its evolution dynamics. In order to understand the behavior of the problem under investigation and develop the initial model, the behavior of the key variables was analyzed and their reference modes were established.

Figure 1 shows the expected behavior over time of the maintenance cost of a software product when influenced by technical debt accumulation. If no should-fix violation occurs, there would be no technical debt. Assuming that the maintenance team size remains constant over time, the maintenance cost would remain stable at an optimum level ("*Optimal maintenance*"). This residual cost occurs because even without technical debt, perfective and corrective maintenance activities are still needed to meet new demands and correct latent defects identified at the operation time.

However, as described by the laws of evolution related to the complexity growth and decreasing quality; violations are an intrinsic part of the software maintenance activities. As the principal component of the technical debt grows, its interest, which is measured by the difference between the optimal maintenance cost and the actual cost ("*Maintenance*"), also grows due to degradation of the productivity of maintenance activities caused by the erosion of the maintainability of the software product.

**Figure 1 – Maintenance costs behavior over time.**

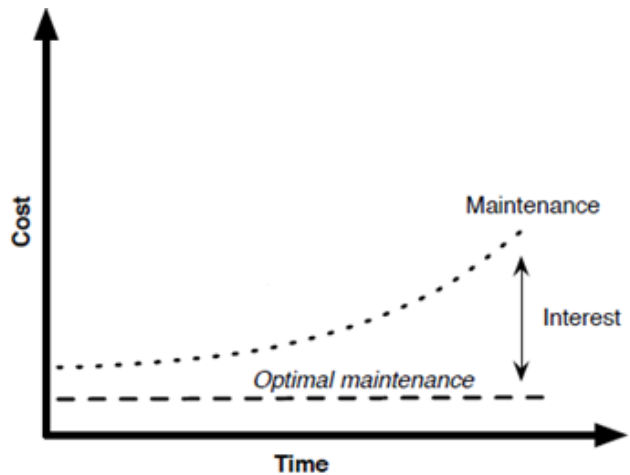

**Source: Adapted from (Nugroho, Visser, & Kuipers, 2011).**

The maintenance cost is defined by the effort required to deliver the demands related to functional and non-functional requirements, which is in turn defined by the product of the maintenance team size and productivity. The maintenance team productivity is affected by should-fix violations accumulation that causes the technical debt principal component to grow which, in turn, reduces the software product maintainability.

Regarding the variables related to the total maintenance effort and the technical debt accumulation, assuming a constant maintenance team size over time, the effort employed throughout each interval ("*ΔEffort/Δt*") is constant. Keeping the maintenance effort constant over time, both the accumulation of the maintenance effort ("*Total effort*") and the technical debt principal accumulation ("*TD principal*") show a linear growth. This behavior is shown in Figure 2.

**Figure 2 – Technical debt principal and maintenance effort behavior over time.**

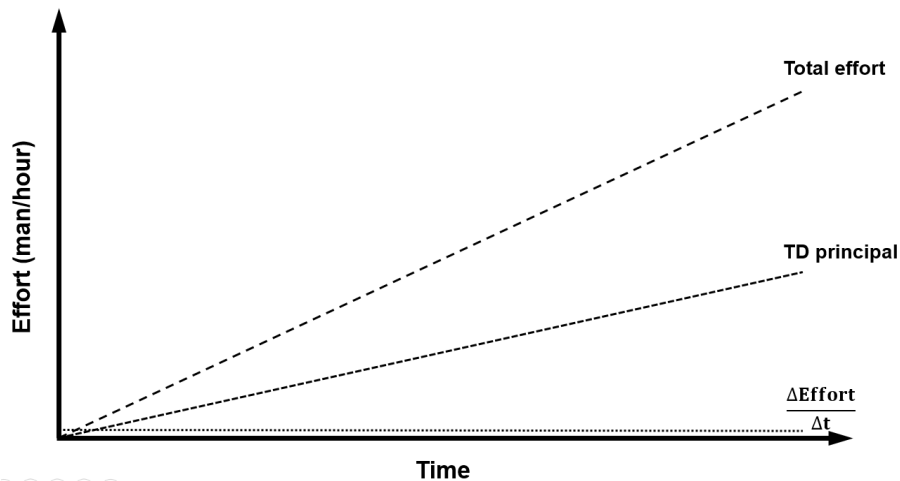

The construction of the model's reference mode regarding the software maintainability, assumes that the maintainability (M) and the effort applied to the perfective maintenance activities (C) have an exponential relationship between them (Bakota et al., 2012), as described in equation:

$$M(t) = e^{-\lambda C(t)}$$

The maintainability proposed by Bakota et al. (2012) measures the software's disorder that is measured by a scalar value ranging from 1 to 0 (higher value is better). The constant "λ" is the quotient of a maintainability erosion factor and a conversion constant between different units of measurement.

Thus, when applying a constant maintenance effort in each time interval, reproducing the linear growth behavior depicted in Figure 2, the maintainability of the software product shows the exponential decay behavior presented in Figure 3 ("*Real maintainability*").

**Figure 3 – Software product maintainability behavior over time.**

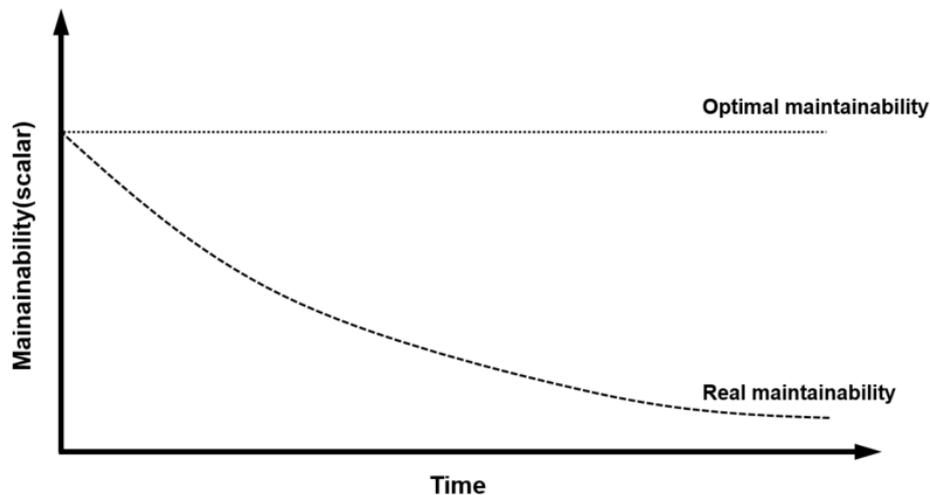

**Source: adapted from (Bakota et al., 2012).**

The formulation of Bakota et al. (2012) does not take into consideration the concept of technical debt management and technical debt payment strategies to restore software product maintainability; issues that are incorporated in the model formulation presented in this paper.

The subsystems diagram, shown in Figure 4, presents an overview of the proposed model, which is composed of four subsystems: "*Software maintenance*", "*Resources management*", "*Software operation*" and "*Software product*". In addition, the subsystem diagram presents the main interactions between these subsystems.

**Figure 4 – Proposed model subsystem diagram.**

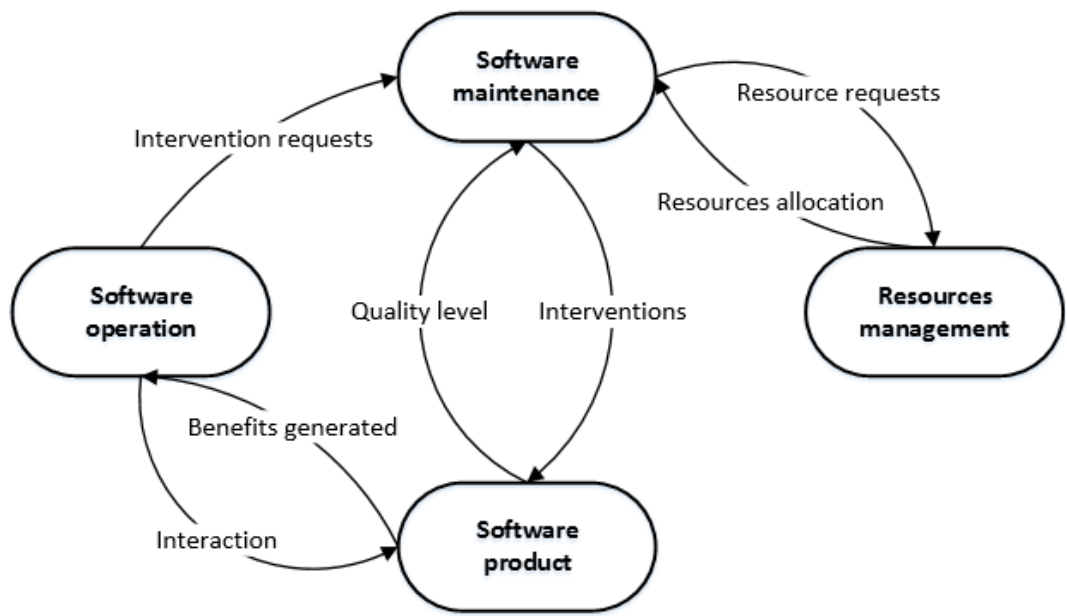

The subsystem diagram is also important to define the boundaries of the model, explicitly stating its structures and assumptions. Table 2 provides a brief description of each of the model's subsystems.

**Table 2 – Proposed model subsystems descriptions.**

| Model subsystem | Description |
| --- | --- |
| **Software product** | It represents the software product, which contains features related to functional and non-functional requirements and interacts with the external environment throughout its life cycle (users and operating context). The laws of software evolution also influence the software product, such as continuing changes; decreasing quality; increasing complexity and continuing growth. In the model, they are characterized by a set of functionalities available and the technical debt incurred due to the violations (intentional or not) over the maintenance performed. |
| **Software operation** | Corresponds to the various users from the business areas of an organization that interacts with the software product. Based on the functionalities available, formulate their perceptions related to perceived ease of use and perceived usefulness, use the system and establish the level of satisfaction. From the provided functionality and system quality, require interventions related to perfective maintenance. |
| **Software maintenance** | It includes elements related to the execution of software maintenance activities performed to deliver the demands received from its users (represented by an activity backlog). In addition, this subsystem is also responsible for performing preventive maintenance activities to reduce the level of technical debt and preserving the software product maintainability. |
| **Resources management** | It represents the investment of resources (financial, personnel etc.) for performing activities related to perfective and preventive maintenance. The availability of the maintenance team is represented as a finite resource that imposes restrictions on the ability to perform the interventions necessaries and its allocation constitute a trade-off analysis between reducing the technical debt (preventive maintenance) and meeting the demands of its users for functionalities (perfective maintenance). |

Table 3 shows the model boundary chart with the variables considered endogenous, exogenous or even excluded.

**Table 3 – Proposed model boundary chart.**

| Endogenous | Exogenous | Excluded |
| --- | --- | --- |
| Maintenance backlog | Software product growth rate | Corrective and adaptive maintenance activities |
| Production library (functional requirements available) | Refactoring effort necessary | |
| | Refactoring overhead | |

| | |
|---|---|
| Technical debt | Software infrastructure (hardware, operation system, network etc.). |
| Software product maintainability | |
| Perfective and preventive maintenance activities | Support services (training, helpdesk etc.) |
| Maintenance team | User satisfaction levels |
| Maintenance team productivity | Data quality model |
| Allocation policy | |

Figure 5 presents the causal loop diagram, identifying the causal relationships between elements of the proposed model, how they affect each other and constitute an important tool for representing its feedback loops. The diagram consists of nodes (elements) and their relationships (arrows). Those relationships can be positive or negative (indicated by the corresponding signal symbol at the end of the arrow). Each closed loop represents a balancing ("B") or reinforcing ("R") behavior, depending on the set of polarities of the relationships involved.

**Figure 5 – Causal loop diagram.**

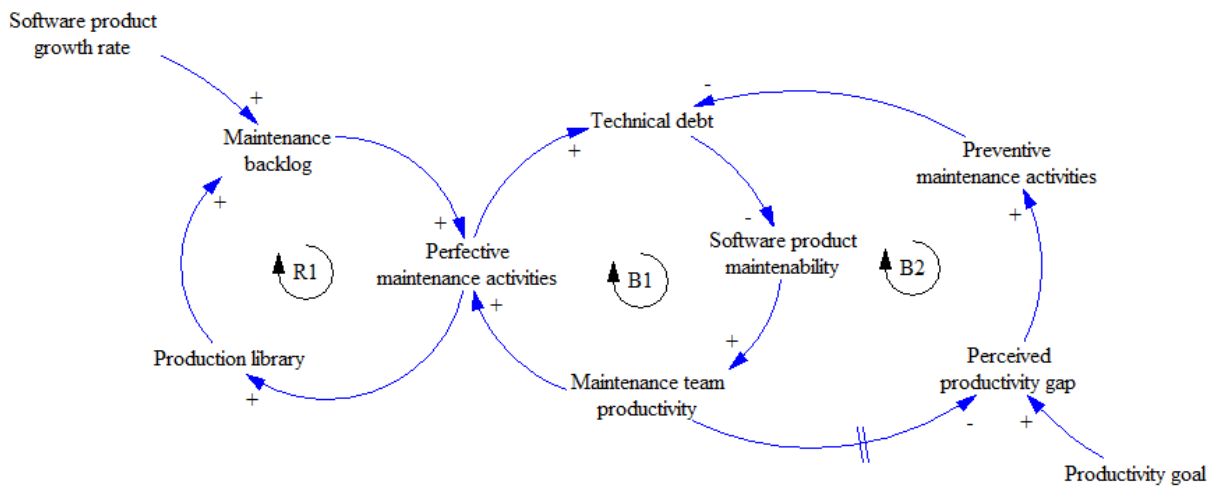

To understand the effect of the technical debt accumulation on the software maintenance productivity and the resource allocation policies, the archetype "Growth and Underinvestment" (Senge, 2006) was used. This archetype consists of three feedback loops, one reinforcing and two balancing (one of them with a delay).

> "*Growth approaches a limit which can be eliminated or pushed into the future if the firm, or individual, invest in additional 'capacity'. But the investment must be aggressive and sufficiently rapid to forestall reduced growth, or else it will never get made. Oftentimes, key goals or performance standards are lowered to justify underinvestment. When this happens, there is a self-fulfilling prophecy where lower goals lead to lower expectations, which are then borne out by the poor performance caused by underinvestment*." (Senge, 2006, p. 399)

The feedback loops depicted in Figure 5 are also characterized by some of the "Laws of Software Evolution" (M. Lehman & Ramil, 2006). The reinforcing loop "*R1*" represents the laws of continuing growth and change, which state that a software product in operation must be continually adapted and enhanced; else it becomes progressively less satisfactory in use over its lifetime. The first balancing loop "*B1*" represents the law of "increasing complexity" that says that as a software product is changed during its lifetime, its structural complexity increases (violations are incurred and technical debt principal build up). As the software product complexity increases, it becomes more difficult to adapt and evolve (reduces maintainability and team productivity), unless work is done to maintain or reduce its complexity (represented by the second balancing loop "*B2*").

## 5 MODEL FORMULATION

The proposed model was built and simulated using the software AnyLogic PLE version 7.2.0 (AnyLogic, 2015). Figure 6 shows the stock and flow diagram that contains the three feedback loops described in the causal loop diagram (Figure 5).

**Figure 6 – Stock and flow diagram.**

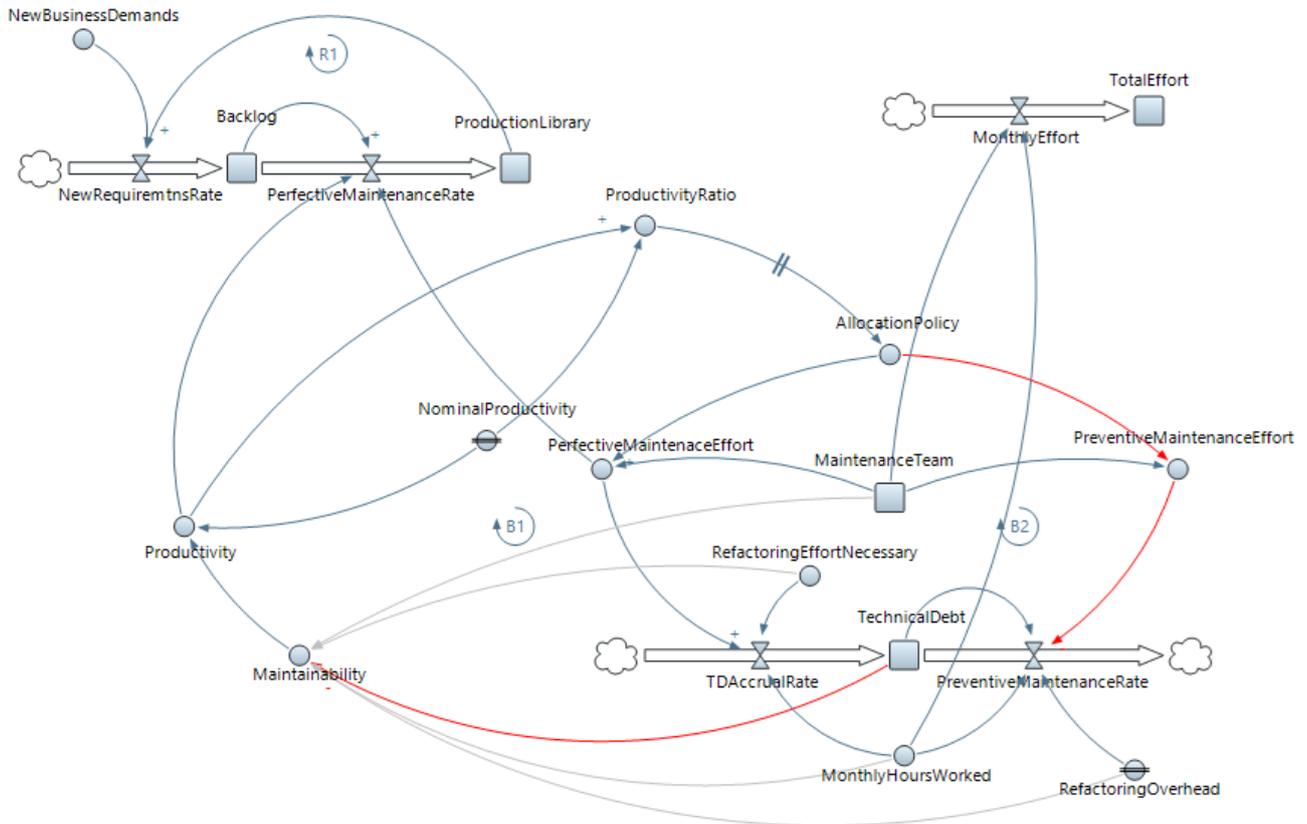

There are four stocks in the model (state variables): "*Backlog*" (functional requirements accumulated and waiting for development); "*ProductionLibrary*" (size of the software in operation, measured in function points); "*TechnicalDebt*" (technical debt principal, measured in man-hour that represents the total effort necessary to remove all the violations accumulated); and "*MaintenanceTeam*" (team size that was assumed constant over the time horizon simulated).

The model's relationships, variables, and equations were developed based on an extensive literature review and secondary data, and are shown in Table 4.

**Table 4 – Proposed model's formulas and units.**

| Formulation and comments | Units |
|---|---|
| $NewBusinessDemands = 0.07$ | *1/year* |
| The fraction of new requirements generated from the size of the software product in operation. | |
| $NewRequirementsRate(t) = ProductionLibrary(t) \cdot NewBusinessDemands$ | *function point/ month* |
| The rate of new functional requirements generated. | |

$$Backlog(t) = Backlog(0) + \int_0^t NewRequirementsRate(s) \cdot ds \qquad \text{function point}$$

The stock of functional requirements that constitutes the total amount of work pending for perfective maintenance activities.

$$ProductionLibrary(t) \\ = ProductionLibrary(0) \\ + \int_0^t PerfectiveMaintenanceRate(s) \cdot ds \qquad \text{function point}$$

The stock represents the size of the software product in operation and measures the amount of function points developed and deployed for use.

$NominalProductivity = 4.65$     *function point/person/month*

The nominal productivity of each person of the maintenance team.

$MonthlyHoursWorked = 160$     *man hour/month*

The amount of hours per month that a member of the maintenance team effectively devotes for performing maintenance activities.

$$Maintainability(t) = e^{\left(\frac{RefactoringOverhead \cdot TechnicalDebt(t)}{TimeHorizon \cdot MaintenanceTeam \cdot MonthlyHoursWorked \cdot RefactoringEffortNecessary}\right)} \qquad \text{dimensionless}$$

Represents the software's disorder and is measured by a scalar value ranging from 1 to 0 (higher value is better).

$Productivity(t) = NominalProductivity \cdot Maintainability(t)$     *function point/person/month*

The effective productivity of each person of the maintenance team, which is affected by the software product maintainability.

$$PerfectiveMaintenanceEffort(t) \\ = MaintenanceTeam \cdot MonthlyHoursWorked \cdot AllocationPolicy \qquad \text{man hour/month}$$

Total effort devoted to performing perfective maintenance activities.

$$PerfectiveMaintenanceRate(t) \\ = \min(PerfectiveMaintenanceEffort \cdot Productivity, Backlog) \qquad \text{function point/month}$$

Actual rate of perfective maintenance for delivering new product functionality to production library.

$$PreventiveMaintenanceEffort(t) \\ = MaintenanceTeam \cdot MonthlyHoursWorked \\ \cdot (1 - AllocationPolicy) \qquad \text{man hour/month}$$

Total effort per month devoted to performing preventive maintenance activities.

$$PreventiveMaintenanceRate(t) \\ = \min(PreventiveMaintenanceEffort \cdot Productivity, \\ TechnicalDebt) \qquad \text{man hour/month}$$

Actual rate of preventive maintenance for paying technical debt and reduce its principal.

$RefactoringEffortNecessary = 0.3$     *dimensionless*

Represents the fraction of the effort devoted to the perfective maintenance that would be necessary for refactoring the software product code to clean up the should-fix violations.

$RefactoringOverhead = 2$     *dimensionless*

The ratio of overhead activities (testing, documenting etc.) for delivering refactoring and regular development code in production. Usually, when refactoring it is necessary to test larger portion of the code base than new developments.

$$TotalEffort(t) = TotalEffort(0) \qquad \qquad man\ hour$$
$$+ \int_0^t MaintenanceTeam(s) \cdot MonthlyHoursWorked \cdot ds$$

Stock representing the total effort devoted to maintenance activities, including perfective and preventive.

The variable "*RefactoringEffortNecessary*" represents the percent of effort spent by developers for refactoring the actual software system to clean up not optimal code base. Many developers usually mix the development and refactoring activities, and therefore, it is difficult to precisely measure the actual amount of time. In the proposed model, we used 30% of the total monthly effort (Cao, Ramesh, & Abdel-Hamid, 2010).

The "*AllocationPolicy*" variable represents the fraction of the total effort available of the maintenance team that will be allocated for the perfective maintenance, the remaining effort will be allocated for preventive maintenance. The formula of this variable represents the resource allocation policies, which are discussed in the next section.

## 6 RESULTS AND DISCUSSION

In the next subsections, two scenarios were analyzed to evaluate the impact of different resource allocation policies for maintenance activities. The time horizon used was eleven years. This period represents the regular lifetime for Management Information Systems (MIS) of the size of 10.000 function points (Jones, 2008).

The initial conditions of the variable used for simulating the proposed model are presented in Table 5.

**Table 5 – Initial conditions of the proposed software maintenance model.**

| Element model | Initial values |
| --- | --- |
| *ProductionLibrary* | 10.000 function points |
| *MaintenanceTeam* | 14 persons |
| *Backlog* | 0 function point |
| *Maintainability* | 1 dimensionless |
| *AllocationPolicy* | 1 dimensionless |
| *TotalEffort* | 0 man-hour |

### 6.1 Perfective maintenance only

The first simulated scenario was to set the variable "*AllocationPolicy*" to a fixed value of "1", meaning that all the resources available were allocated for perfective maintenance. This policy intends to reduce the "*Backlog*" and deliver new business demands, increasing the size of the software product in operation. Figure 7 show the behavior over time of the nine key variables of the model: "*ProductionLibrary*"; "*ProductivityRatio*" (the ratio between the actual and nominal maintenance team productivity); "*AllocationPolicy*" (the percentage of the maintenance team effort devoted to perfective maintenance); "*Maintainability*"; "*Backlog*"; "*TotalEffort*"; "*ΔEffort/Δt*"; and "*TechnicalDebt*".

The top-left curve shows the growth of the size of the software product in operation ("*ProductionLibrary*"). The curve resembles an asymptote shape due to the maintainability and maintenance productivity exponential decay shown in the top-right graphic. As the productivity fall, the delivery of new functionality slows down and the requirements "*Backlog*" exhibit an exponential grow. The "*TechnicalDebt*" accumulation, bottom-right graphic, present a linear growth because of design choice made during the model formulation (constant maintenance team size), and fixed values for "*AllocationPolicy*" and "*ReffactoringEffortNecessary*" variables.

**Figure 7 – Key variables behaviors over time of perfective maintenance only.**

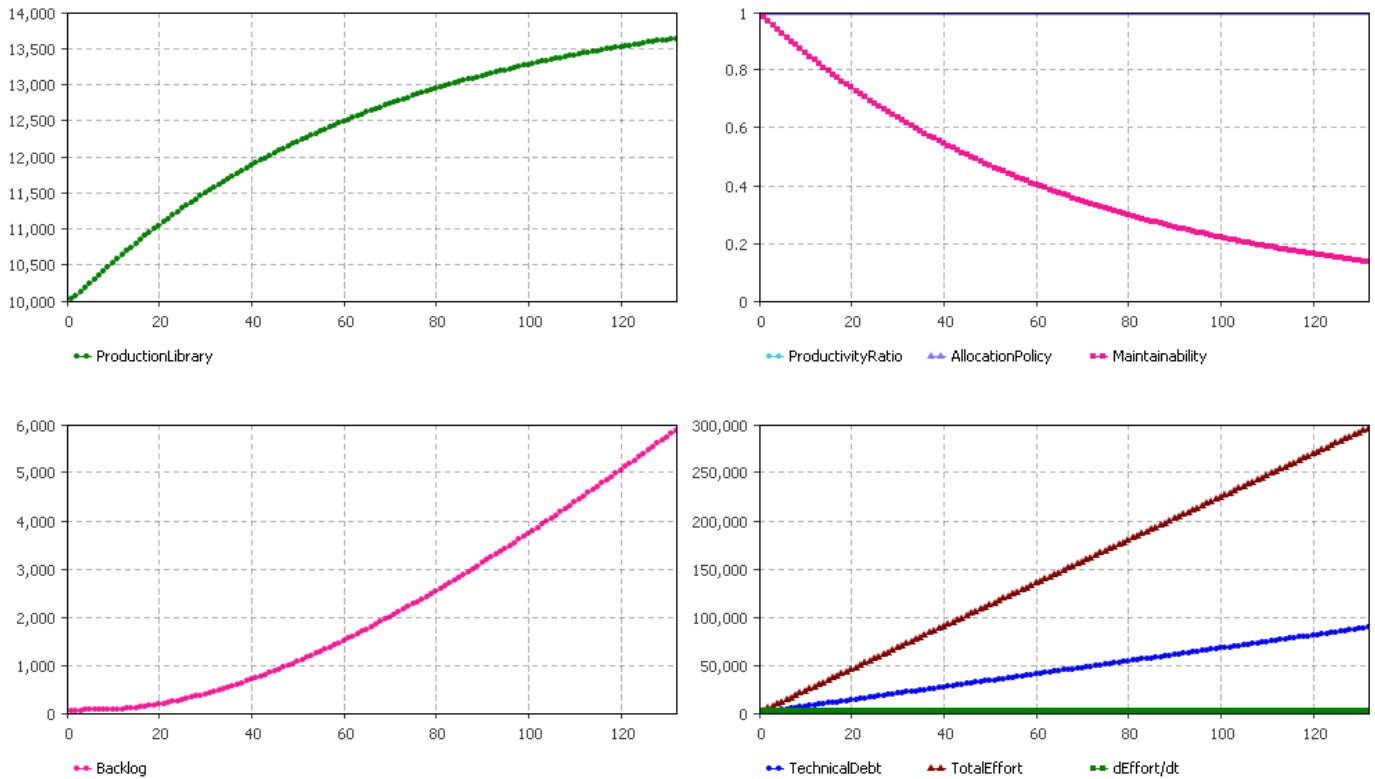

This first scenario represents the hypothesis that was used for formulating the reference modes of the key variables described in section "4 Dynamic Hypothesis". The bottom-right graphic of Figure 7 reproduces the maintenance effort and technical debt principal growth behaviors previously depicted in Figure 2, while the top-right graphic of Figure 7 reproduces the software product maintainability behavior previously depicted in Figure 3.

*6.2   Preventive maintenance trigger due to productivity decay*

The second scenario simulated represents a policy that tries to maintain the "*TechnicalDebt*", incurred during the maintenance of the software product, at some sustainable level that does not cause the productivity of the "*MaintenanceTeam*" to degrade to the point where the cost of maintenance rises to an unfeasible level. In this way, the variable "*AllocationPolicy*" was setup as a table function, according to the graphic depicted in Figure 8.

**Figure 8 – Table function of allocation policy due to productivity decay.**

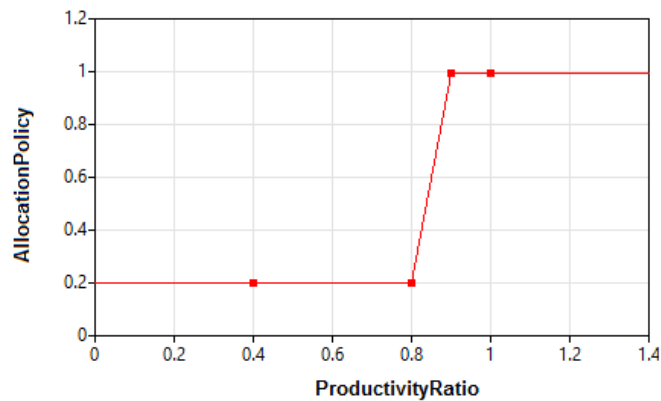

When the "*ProductivityRatio*" is "1", i.e. there no loss of maintainability and productivity, all resources are allocated to perform perfective maintenance. When the productivity starts to fall, and as the "Growth and Underinvestment" archetype states, the investment in capacity needs to be aggressive and quick. Therefore, the allocation quickly changes from "1" to "0.2", meaning that 80% of the resources available change its focus to perform preventive maintenance activities, and only 20% remains focused on the perfective maintenance.

Figure 9 shows the behavior over time of the same nine key variables described in the previous scenario.

**Figure 9 – Key variables behaviors over time of preventive maintenance trigger due to productivity decay.**

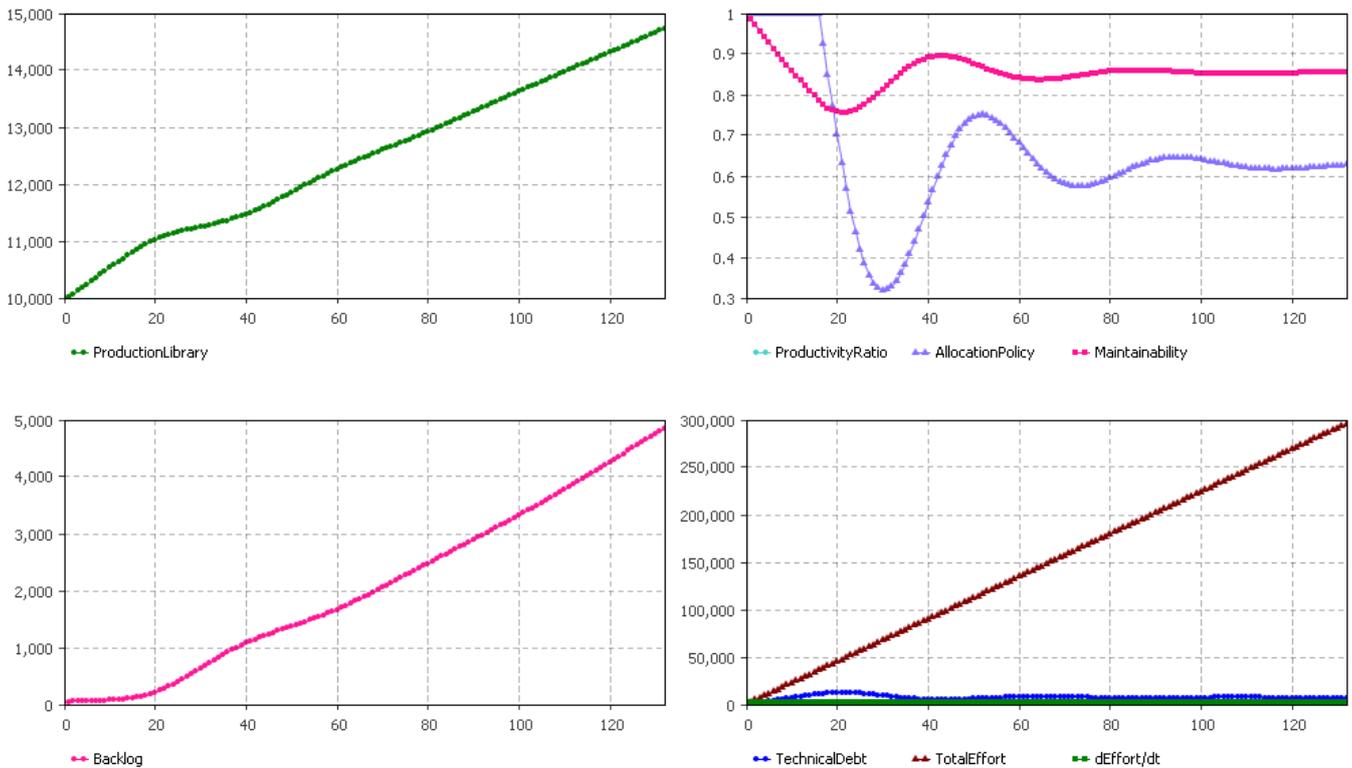

When introducing a non-linear decision rule for the "*AllocationPolicy*" variable (Figure 8), the observed behavior of the key variables dramatically changed. With this new decision rule, when de "*ProductivityRatio*" fall below "1", the "*AllocationPolicy*" start to respond to the productivity changes with a first-order delay of twelve months, interval adopted for an organization to identify the productivity decay and adopt countermeasures (top-right graphic).

When the "*AllocationPolicy*" starts to fall, thus increasing the amount of resources allocated for preventive maintenance, "*TechnicalDebt*" also starts to fall (bottom-right graphic) and the "*Maintainability*" begins to rise (top-right graphic). The opposite behavior is observed when the "*ProductivitryRatio*" rises, in response to the productivity recovery, and the "*AllocationPolicy*" changes again and starts to react with a delay to focus on the perfective maintenance, slowing down the "*Backlog*" growth (botton.-left graphic).

## 6.3 Discussion

The distinct dynamic behaviors observed in the simulation of the two scenarios described in the previous subsections occurs due to the dominance shifts of the feedback loops shown in the causal loop diagrams (Figure 5).

At the beginning of the simulation of the first scenario ("*Perfective maintenance only*"), the reinforcing loop "*R1*" dominates and the software product grows fast, while the should-fix violations accumulate and the technical debt principal grows. As the time passes, the balancing loop "*B1*" starts to dominate and the speed of the software product growth slows down and the backlog starts to build up. In this scenario, the second balancing loop "*B2*" does not get activated since the "*AllocationPolicy*" is fixed at value "1".

In the simulation of the second scenario ("*Preventive maintenance trigger due to productivity decay*"), the reinforcing loop "*R1*" also dominates at the beginning. However, just before the technical debt principal accumulates to a level where the balancing loop "*B1*" starts to dominate the dynamics of the system, the "*AllocationPolicy*" shifts the dominance to the balancing loop "*B2*", focusing on paying technical debt principal and restoring maintenance productivity. When it reaches its goal, the "*AllocationPolicy*" rule changes the dominance again to the reinforcing loop "*R1*" until an equilibrium is almost reached ("*AllocationPolicy*" around 0.85).

## 7 CONCLUSION

This work presented a model built using the System Dynamics approach that can be used for exploring the long-term effects of different resource allocation policies for software maintenance process for managing technical debt accumulation over time. Different strategies demonstrated different behaviors of key variables, such as software product size, backlog, maintenance productivity, maintainability and technical debt principal.

When comparing the "Perfective maintenance only" and the "Preventive maintenance trigger due to productivity decay" policies, it was possible to notice that even when transferring some resources from perfective maintenance to preventive maintenance (for paying the technical debt principal), the second policy ended up with more functionalities deployed, lower backlog, higher productivity ratio and lower technical debt accumulated.

Although some promising results were obtained, the model still has limitations that should be addressed in future works. The corrective and adaptive maintenance activities also require resources over the software operation, the maintenance team that was kept constant changes over time and its members have different characteristics (learning curve, experience, productivity, time for training, communication overhead etc.).

Additionally, the model must be submitted to a rigorous set of tests, sensitive analysis and validations to increase its confidence. Other scenarios should also be analyzed to formulate and evaluate the impact of different intervention policies in the operation and maintenance phases that optimize the trade-off between maintenance costs, software product capacity do adapt and evolve and delivering business needs.

ACKNOWLEDGMENT

We would like to thank the "Coordenação de Aperfeiçoamento de Pessoal de Nível Superior (CAPES)" and the "Conselho Nacional de Pesquisa e Desenvolvimento (CNPq)" for providing research fellowships.